\documentclass{cup-hpl}
\usepackage{color,soul}
\usepackage{lipsum}
\usepackage{natbib}
\usepackage{comment}
\DeclareUnicodeCharacter{2009}{\,}

\begin{document}

\newtheorem{theorem}{Theorem}

\shorttitle{PIS}                                   
\shortauthor{L. Volpe}

\title{A Platform for Ultra-fast Proton Probing of Matter in Extreme Conditions}

\author[1,2]{Luca Volpe}
\author[1]{Alberto Perez}
\author[1]{Teresa Cebriano Ramírez\corresp{Address of corresp.\email{tcebriano@clpu.es}}}
\author[3]{Carlos Sánchez Sánchez}
\author[4]{Sofia Malko}
\author[1]{Alessandro Curcio}
\author[3]{Berkhahoum Kebladj}
\author[3]{Samia Khetari}

\address[1]{Centro de Laseres Pulsados, Building M5, Science Park, Calle Adaja 8, 37185 Villamayor, Salamanca, Spain}
\address[2]{ETSI Aeronáutica y del Espacio, Universidad Politécnica de Madrid, 28040 Madrid, Spain}
\address[3]{Department of Fundamental Physics, University of Salamanca, 37008 Salamanca, Spain}
\address[4]{Princeton Plasma Physics Laboratory, 100 Stellarator Road, Princeton, NJ 08536, USA.}

\begin{abstract}
Recent developments of Ultra Short and Intense laser system paved the way for the generation of short and brilliant proton sources that can be used to study plasmas at extreme conditions in the context of High Energy Density Physics. Energy selection, focusing and transport of such sources is still a challenge that is not completely solved. Here we present a novel and simple design for a isochrone magnet capable of angular and energy selection of proton sources by reducing the temporal spread. The isochrone selector is demanded for studying proton stopping power in Warm Dense Matter close to the Bragg peak (proton energy below 1 MeV) where the reduced investigated masses imply a short plasma stagnation time ($<$ 100 ps). The proposed selector can also be used at higher proton energy by pushing the final time spread at the level of few ps. Analytical estimations and Monte Carlo simulations are presented to validate the proposed configurations. 
\end{abstract}

\maketitle

\section{Introduction}
The advent of PW-class CEP laser systems in the last decades broke ground for the acceleration of ultra-short and brilliant proton bunches. The latter have been used to probe matter in extreme conditions such as the Warm Dense Matter generated via isochoric heating (via laser, X-ray, electrons and ions) which is relevant for of High Energy Density Physics, laboratory Astrophysics, Planetology and, especially, for Laser fusion \cite{hur19, zyl22, mar18, rot01}. High energy density (HED)  plasmas, including Warm Dense Matter (WDM), are important and fascinating states of matter that can be now easily generated by lasers or by laser-driven secondary sources such as X-rays \cite{Zyl15}, electrons or ions \cite{patel2003}. The sub-ps dynamics of such states is important to understand the mechanism of the electron-electron and electron-ion thermalisation far from the equilibrium. Laser-Driven Proton Sources (LDPS) can be used to study Ion Stopping Power (ISP) in HED and WDM as well as to be used for radiography \cite{mac06,vol11} and/or deflectometry \cite{Li08}. Ultra-short Laser-driven proton bunches are becoming relevant also in the context of biology and medical physics where the effects of fast dose deposition in tissues is now under investigation. In addition, recently emerging experimental stations offer combinations of intense solid-state lasers with the 4$^{th}$ generation of X-ray Free Electron Lasers \cite{Zas21} sources, that are capable to provide highly brilliant photon sources in the keV region. A new generation of experiments is now possible, where fast-heated plasma states can be investigated by using fast laser-driven electron or proton sources. Highly brilliant proton beams are routinely generated via Target Normal Sheath Acceleration (TNSA) \cite{wilks2000} mechanism with exponential energy spectrum ranging up to few tens of MeV's and duration of the order of few ps. Numerical simulations suggest the possibility of improvements in the quality of such sources in terms of maximum energy and energy spread. This can be accomplished by increasing the laser intensity as well as improving the pulse temporal contrast and the beam spatial quality. Conventional accelerators can provide proton sources with high  brightness, low energy spread and very small divergence. However, LDPS are still relevant due to some peculiar characteristics like: i) ultra-short pulse duration ($\sim $  ps) ii) High Repetition Rate (HRR) capability (up to 1 KHz) and iii) Small size i.e Table Top Machines (Few meters dimension and low cost).
Indeed, LDPS are extremely short \cite{dro16}, being able to provide an almost instantaneous ($<$ ps) dose, not comparable with other proton sources. Such dose can be comparable with standard doses in HRR operation. Finally, the reduced dimension of laser systems compared with standard accelerators are improving day by day. Indeed, laser companies are now able to provide compact hundreds of TW scale laser systems occupying a few meters area. These systems can work at 10s of shots per second and technological development is promisingly increasing the HRR up to 100 shots per second. Smaller laser systems can reach tens of TW with higher repetition rates up to 1 kHz, see for example the Silos system of ELI-ALPS (7 fs, 30 mJ, 1 KHz).
LDPS can be used to investigate plasma states  at unprecedented short time scales. Reducing the pulse duration of the “driver” below the hydrodynamic time scale of matter (i.e. below 1 nanosecond) permits to heat the matter isochorically; at the same time acceleration of short charged particle beams \cite{Alb21} with lasers offers the possibility to perform “pump and probe” experiments in which the same laser is used to generate the plasma sample and accelerate the proton probe. However, many of these experiments require quasi monoenergetic and short proton pulses while TNSA mechanism generates proton beams with a broadband energy spectrum. Therefore additional beam manipulations, such as energy selection or spatial focusing is needed. Energy selection and beam collimation can be done by using special configuration of permanent magnets: dipoles, quadrupoles, etc. 
Unfortunately, such manipulation has a cost in terms of final proton pulse duration that increase proportionally to the size of the magnetic device. In this line of research several developments are already underway. The most relevant examples are  ELI-MAIA ion transport system at Eli beamline in Czech Republic \cite{mar18,cir20} and the pair of pulsed solenoids established at Draco system at HZDR in Dresden Germany \cite{kro22}. Other smaller-size systems were also developed for dedicated experiments \cite{che14,ten16}. All the above-mentioned approaches can provide MeV-energy proton beams with time duration of the order of $\mu$- seconds or in the best case can go down to hundreds of ns. For several applications this is not representing a limitation. However, as stated before, the new opened experimental possibilities, by using $\mu$m-size and ps-time plasmas, require shorter proton/ion bunches down to few picoseconds and, for biological applications, is interesting to go even further down to sub-picosecond time duration to investigate coherent absorption processes. The ion time spread due to transportation is more important for proton energies below the MeV level, for example a proton beam with central energy of  1 MeV and energy spread of 100 keV increase the temporal length at an approximated rate of 70 ps/cm.
Recently, at the Centro de Laseres Pulsados (CLPU) an experimental campaign has been conducted to measure proton stopping power in Warm Dense Plasmas. In such experiment the Ti:sa 30 fs 200 TW systems at CLPU was used to both generate the WDM and the proton beam whose energy loss was measured after passing through the laser-driven WDM sample \cite{Mal22}. The novel experimental approach has improved time and energy resolution guiding new experimental activities focused on micrometric size and picosecond and sub-picosecond time plasmas with implications in several interesting fields of science among which Inertial Confinement Fusion (ICF) \cite{hur19,zyl22} and ion-driven fast ignition \cite{rot01,Fer14}. A precise knowledge of ion stopping in WDM is also essential for understanding proton transport in matter \cite{kim15} and proton isochoric heating \cite{patel2003,mcg20}. Such experiments have applications for studying the structure \cite{man10}, the equation-of-state \cite{pin19} and the transport properties of dense plasmas \cite{gra20}, like conductivity \cite{pin19,mck17} and thermal equilibrium \cite{whi12} of WDM samples. Other applications include plasma diagnostics using ion beams \cite{mac06,vol11}. Developing ultra-short proton pulses was mainly driven from Ion stopping power in WDM experiments even if there are at least two more important applications of such pulses.  The first is Proton Fast Ignition approach to Inertial fusion \cite{rot01} where the requested duration of the proton beams falls below 20 ps. This can be reached only by developing a complex time-compressor transport line \cite{he15}. The second application is related to the use of short proton pulses in the context of the “fast dose deposition” \cite{ras21} approach to cancer therapy in Medicine, that is an emerging field of investigation.  
In the above mentioned experiment \cite{Mal22} a compact magnetic selector \cite{api21} has been used to manipulate the energy-space properties of a laser-driven sources sent to a laser generated WDM sample. The final temporal duration of the selected proton beam was about 400 ps (@ the WDM sample) a bit longer than the characteristic WDM stagnation time  with the central energy of the proton beam fixed to 500 keV $\pm$ 25 keV. The magnetic device was 6 cm long and the total distance between proton source and the WDM sample was around 8 cm. The compact magnetic selector was made by a dipole magnet with a system of pinholes and slits at the entrance to control the proton divergence, and at the exit to control the energy band that could be reduced at the expense of the total number of selected protons. A new and optimised version of the selector can help to reduce the temporal stretch but going below 100 ps requires a different approach.

Here, we proposed a simple way to perform a consistent reduction of the proton pulse duration, by using an isochrone ion selector, down to few 10s of ps @ energies $<$ 1 MeV. The selected proton pulses could be used for experiments that require high temporal and spectral resolution, like ultrashort pump-probe experiments and proton isochoric heating. Another relevant application is certainly proton/ion stopping power studies in WDM  with an unprecedent energy resolution very close to the Bragg peak region (region where the proton energy is comparable with the thermal energy of the WDM electrons) where most of the theoretical predictions are in contradiction. Ultrashort LDPS will open a new horizon for applications in the field of biology and medical physics for the study of fast dose deposition in proton therapy. 
\section{Design}
\begin{figure}
	\centering
		\includegraphics[width=0.45\textwidth]{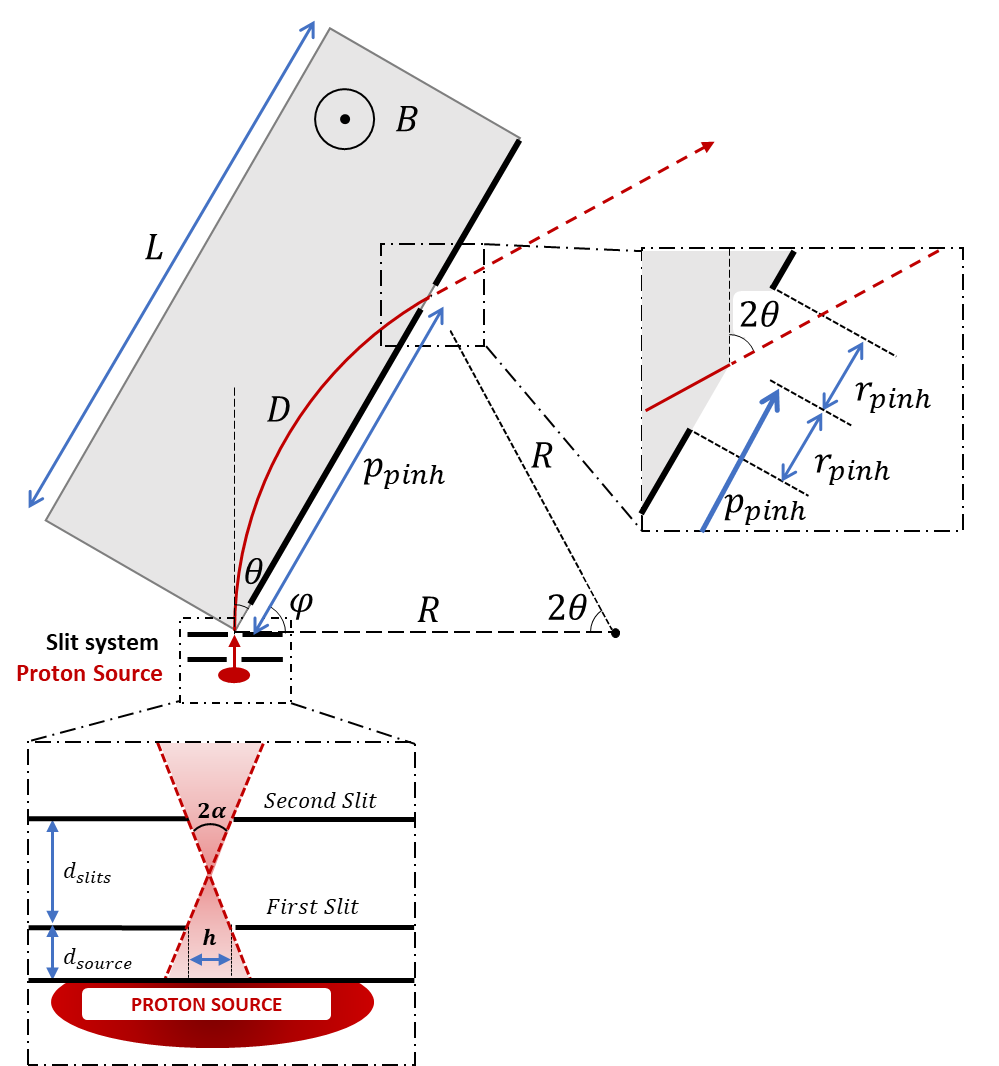}%
	\caption{Scheme of the proposed device. Lower and upper right inset are zoomed pictured of the slit system and the final pinhole respectively. Angles $\theta$ and $\varphi$ are complementary.}
\label{Dipole diagram}
\end{figure}

We assume here that proton sources are generated with the conventional TNSA mechanism with a characteristic broadband energy spectrum and divergence. The magnetic selector Fig. \ref{Dipole diagram} is composed by three main parts: i) a first collimator, ii) an isochrone magnetic transporter and iii) a final energy selector. 
The collimator is composed by two slits (Fig. \ref{Dipole diagram}, lower inset) placed close to the proton source. It can discriminate protons according to their divergence. Although introducing losses, this stage reduces the angular dispersion of the proton pulse, which in turn reduces the temporal dispersion at the exit of the magnet. The collimated protons enter into the magnetic transporter made by a dipole (upper left figure), where a constant magnetic field disperses different proton energies, while maintaining the temporal dispersion. Finally, the energy selector, made by a exchangeable and movable pinhole, allows only the exit of protons whit certain energies (upper right inset). The designed magnetic selector permits to obtain a final collimated proton pulse extremely narrow both in energy and time.
According to Fig.\ref{Dipole diagram} the system has four adjustable parameters: the distance between the slits($d_{slits}$), the angle between the proton pulse and the side of the magnet ($\theta$), and the position($p_{pinh}$) and radius ($r_{pinh}$) of the pinhole. The first parameter changes the temporal dispersion of the output pulse ($\Delta T$), and it is usually set to the minimum. It also changes the angular dispersion at the exit and is often significantly narrow (1 - 3 mrad). The last three parameters control the central energy ($E_c$) and the energy dispersion ($\Delta E$) of the output beam, and they are selected according to experimental requirements.

\subsection{Temporal selection}
The dynamics of a moving charged particle inside a constant magnetic field is ruled by the Lorentz force $\vec{F}_L = q \; (\vec{v} \times \vec{B})$, where $q$ and $\vec{v}$ are the particle charge and velocity respectively, and $\vec{B}$ is the magnetic field. Since Lorentz force is always perpendicular to the speed, it causes particles to describe circular orbits, being the  cyclotron radius $ R_c = m v / q B$, where $m$ is the mass of the particle. Considering non relativistic protons with energy $E_c = m v^2 /2$, the cyclotron radius can be written as:
\begin{equation}
R_c =\frac{\sqrt{2 m}}{q} \frac{\sqrt{E_c}}{B}
\label{Equation: radius}
\end{equation}
As a consequence, at a constant field, the radius of the proton trajectories depends only on the energy of the protons.
If the limits of the dipole have been set as shown in Fig. \ref{Dipole diagram}, the total distance that the proton travels inside the dipole is $D = 2 \theta R_c$, where $\theta$ is the angle between the proton initial speed and the magnet side. The whole Time of Flight (ToF) of this part of the trajectory $t_{dipole}$ is:
\begin{equation}
t_{dipole} = D/v = \frac{2 m }{q B} \theta
\label{Eq: Time of flight}
\end{equation}
Notice that it does not depend on the energy of the protons, but it does on the angle $\theta$. This implies that the only temporal dispersion produced inside the dipole is caused by the angular dispersion and not by the energy dispersion. Therefore, if the entrance angle of the protons is limited to $ \theta \pm \alpha$ (as shown in Fig. \ref{Dipole diagram}, lower insert), the temporal dispersion $\Delta t_{\alpha}$  that affects the proton pulse is:
\begin{equation}
\Delta t_{\alpha} = \frac{4m}{q B} \alpha
\label{Eq: Angular time dispersion}
\end{equation}
To achieve a small temporal dispersion, therefore, a small acceptance angle is critical, and this is the reason why the slits system before the dipole is necessary. However, this system requires that protons travel some distance in free space, with increased temporal dispersion due to the differences in kinetic energies. The distance between the source and the first slit ($d_{source}$) is meant to be as close as practically possible, while the distance between the slits depends on the needed acceptance angle:
\begin{equation}
d_{slits} = \frac{h}{\tan(\alpha)}
\label{Eq: slit distance}
\end{equation}
The time dispersion caused by this First Flight ($\Delta t_{\text{ff}}$), assuming a divergence angle $\alpha$ very small, is:
\begin{equation}
\begin{aligned}
\Delta t_{\text{ff}} = & (d_{slits}(\alpha) + d_{source}) \sqrt{\frac{m}{2}} \times \\
\times &\left[ \frac{1}{\sqrt{E_c - \Delta E})} - \frac{1}{\sqrt{E_c + \Delta E})}\right]
\label{Eq: FF time dispersion}
\end{aligned}
\end{equation}
Those two effects (time dispersion due to angular dispersion $\Delta t_{\alpha}$ and due to first flight $\Delta t_{\text{ff}}$) have opposite dependencies on the acceptance angle. Maintaining the same central energy and energy dispersion, it is possible to change the acceptance angle $\alpha$ by changing the distance between slits $d_{slits}$, but increasing the acceptance angle reduces the temporal dispersion of the first flight and increases the dispersion due to the angular dispersion, and viceversa. The total temporal dispersion due to both combined effects can be described as follow:
\begin{equation}
\Delta t(\alpha, E_c, \Delta E) = \sqrt{\Delta t_{\text{ff}}^2 + \Delta t_{\alpha}^2}
\label{Eq: Total time dispersion}
\end{equation}

\subsection{Energy selection}
\begin{figure}[t]
	\centering
		\includegraphics[width=0.45\textwidth]{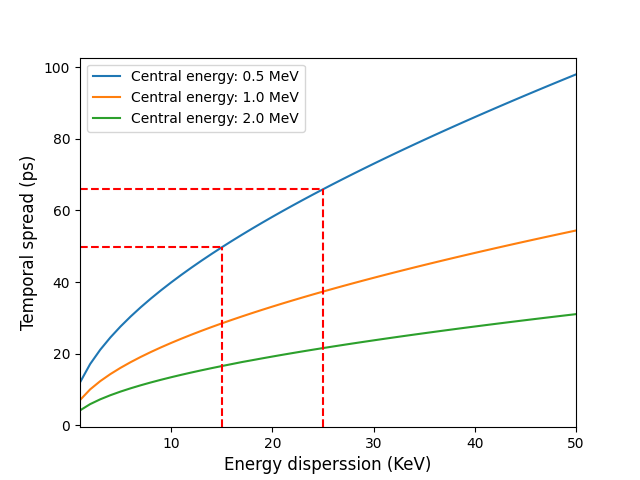}%
	\caption{Time dispersion as a function of the  energy dispersion.for different central energies}
\label{Graphs theory time}
\end{figure}

To control the spectrum of the exiting protons ($E_c \pm \Delta E$), a final pinhole is placed. The expression of the pinholes position ($p_{pinh}$)  and radius ($r_{pinh}$) are:
\begin{align}
p_{pinh} & = k_E \, \sin(\theta) \left(\sqrt{E_c + \Delta E} + \sqrt{E_c - \Delta E}\right) \\
r_{pinh} & = k_E \, \sin(\theta) \left(\sqrt{E_c + \Delta E} - \sqrt{E_c - \Delta E}\right)
\label{Eq: Pinhole coordinates}
\end{align}
where  $k_E = \sqrt{2 m}/(q B)$.
It is also useful to express the same equations with the central energy and energy dispersion as a function of the position and radius of the pinhole so they can be checked during the experiment.
\begin{equation}
E_c  = \frac{(p_{pinh}^2 + r_{pinh}^2)}{4 k_E^2 \sin^2 (\theta)} \; ; \;\;\;
\Delta E  = \frac{p_{pinh} \cdot r_{pinh}}{2 k_E^2 \sin^2 (\theta)}
\label{Eq: Pinhole coordinates reversed}
\end{equation}
Another useful expression is the maximum proton energy that a dipole can operate as a function of the length of the dipole $L$ and the angle $\theta$:
\begin{gather}
E_{MAX} = \frac{L^2}{4 k_E^2 \sin^2 (\theta)}  \\
E_{MAX}[MeV]\approx   \frac{1.2 \times 10^{-5}}{\sin^2 (\theta)} \;B[T] \; L^2[mm] 
 \label{Energy_limit}
\end{gather}
Considering a typical case, a dipole of L = 20 cm, $B$ = 1 T, and $\theta$ = 30$^\circ$, the maximum energy that this device can operates is 1.9 MeV. In theory, any dipole can operate at any energy if the angle $\theta$ is small enough, regardless of its length. In practice, a tiny angle makes the dipole harder to align, so a compromise is recommended.
Assuming a fixed central energy and energy dispersion, it is possible to  numerically calculate the acceptance angle that achieves the smallest temporal dispersion, which typically lies between 0.5 and 2 mrad. The total temporal dispersion achieved on these configurations is also illustrated in Fig. \ref{Graphs theory time}. We can see that the time dispersion decrease as the relation $\Delta E / E_c$ also decreases. To better evaluate the selector performance we can compare the time spread of a proton beam with the same parameters as in \cite{api21}. Targeting energy of 500 keV $\pm$ 25 keV, and assuming a distance between the source and the first slit $d_{source}$ of 3 mm, we can obtain a proton pulse, at the exit of the selector, with a final duration of around 70 ps, which is $\sim$ 6 times smaller than the one obtained with the previous selector ($ \simeq $ 400 ps). The graph in Fig.\ref{Graphs theory time} shows that, for 500 keV central energy, an even shorter proton pulses ($\sim $ 50 ps) can be obtained by reducing the energy bandwidth down to 15 keV (this is possible by reducing the number of selected particle or by reducing the proton source size). Another operation regime, i.e. with sub-picosecond proton pulses ($ <$ 1 ps), can also be obtained by  considering higher proton energy of 15 MeV with few keVs of energy spread, this will open new possibilities related to fast proton imaging and/or deflectometry. According to eq.\ref{Energy_limit}, in order reach 15 MeV of selected energy the selector configuration must be adapted by enlarging the total size L up to 300 mm and by reducing the angle with respect the normal $\theta=15^\circ$.

\subsection{Transmission coefficient}
Energy Selection is directly dependent on the spatial extension of the used pinholes which imply a strong reduction of the total number of the selected protons. To estimate the transmission coefficient we have to take into account three different elements: i) the source size, ii) the solid angle of acceptance, and the central energy. Firstly, we know that the source is not exactly punctual, but a disc with a radius of few hundreds of um (here we use 150 $\mu m$ approximately \cite{Mal22}). This dimension is significantly larger than the slits aperture typically used in this device ($\sim 10 \; \mu m$), so it is necessary to count what portion of the disc actually has some fire angle. To do this we assume that the emission is homogeneous on all the disc surfaces. Secondly, the combination of the slits system and final pinhole only leaves a narrow solid angle of acceptance for the protons, which only allows the protons with very small divergence to pass. It is possible to geometrically calculate this solid angle for a given energy, and later, integrate which proportion of the protons are emitted in this angle. Finally, this acceptance solid angle is different from energy to energy, being zero far from the central energy and maximum at it. To better quantify the transmission coefficient of the device in our range of interest (i.e. $E_c \pm \Delta E$), we have to integrate it over all that range.

\begin{figure}[t]
	\centering
		\includegraphics[width=0.45\textwidth]{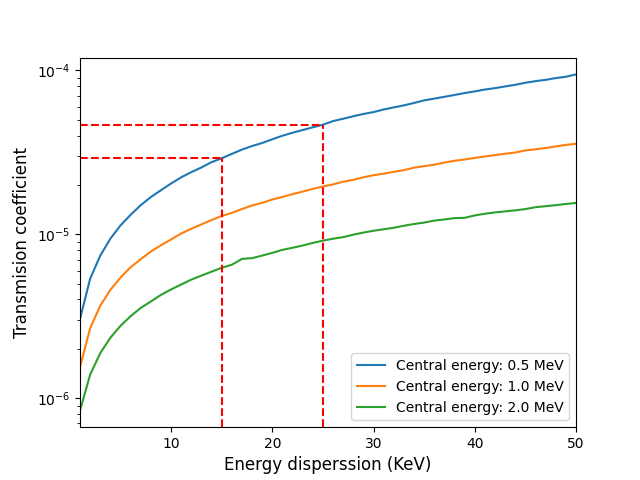}%
	\caption{Time dispersion as a function of the  energy dispersion.for different central energies}
\label{Graphs theory transmission}
\end{figure}

The calculated transmission coefficient is represented as a function of energy dispersion and for different central energies in Fig. \ref{Graphs theory transmission}. Here we can see that, once more, the transmission decrease with the relation $\Delta E / E_c$. It is then essential to make a compromise at the time to design an application, because configuring the device in order to achieve a narrow energy bandwidth implies that the proton flux will be reduced.

\begin{figure}[t]
	\centering
		\includegraphics[width=0.45\textwidth]{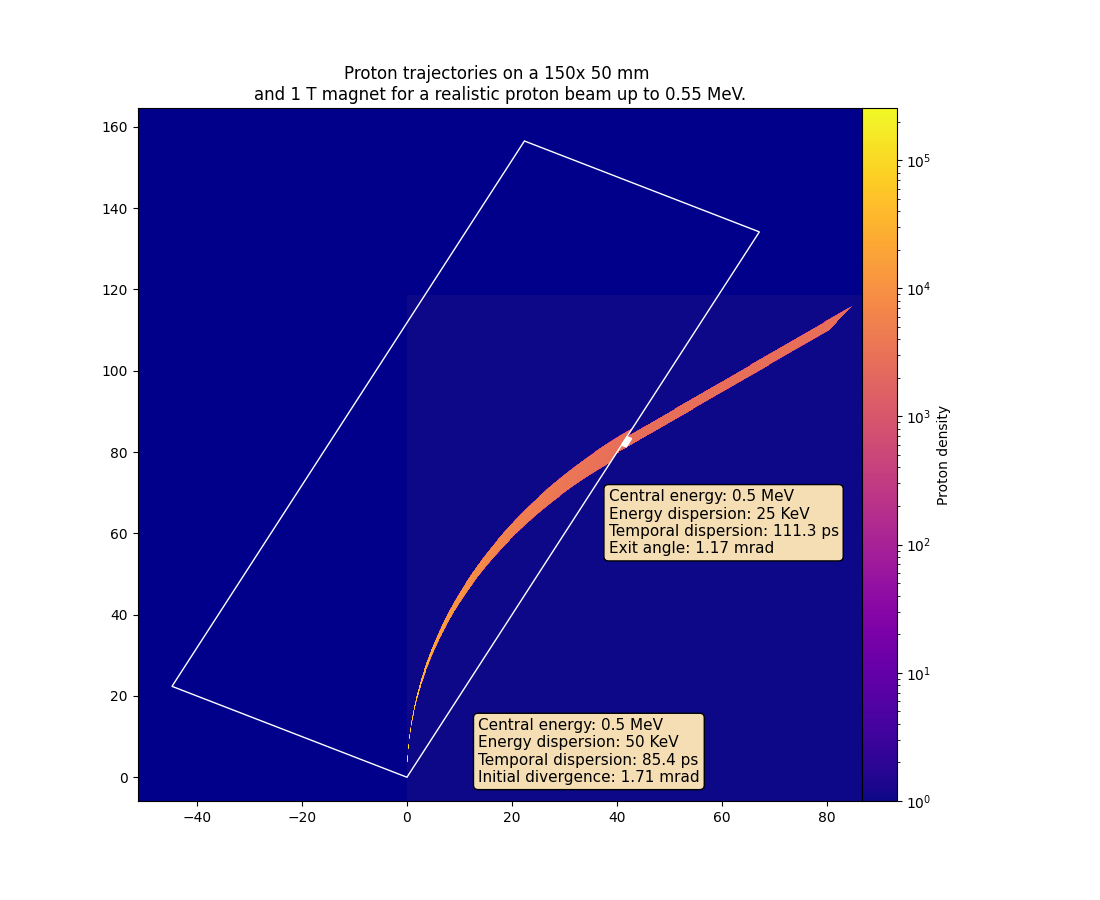}%
	\caption{2D results proton particle transport in the isochrone magnetic selector }
\label{Graph simulation trajectories}
\end{figure}

\section{Simulations}
In order to probe those concepts, a simulation has been carried out by tracking individual protons along its path through the device using a dedicated code. The code has been made using Python programming language and Numpy, Scipy, Numba and Matplotlib libraries. The method used consist of particle tracking of individual protons along its trajectory by using the Lorentz force and Runge-Kutta method.

For the simulation, the distance between the source and the first slit is $d_{source}$ = 2 mm, the distance between slits has been set at 7.6 mm, and the slit space is 8 $\mu m$. The dipole has a magnetic field of B = 1 T, the angle between the dipole side and the proton beam $\theta$ has been set at 63$^\circ$, and the dipole has a length of 15 cm. The pinhole has been placed at a distance of $p_{pinh}$ = 9.14 cm, and has a radius of 2.3 mm, aimed to the energy range of 1 MeV $\pm$ 50 keV. The input proton beam has the typical TNSA angular and energetic distribution of a laser pulse hitting an Al foil target. The interaction area has been estimated as a  75 $\mu m$ radius circular source and 30$^\circ$ FWHM angular distribution. Only protons with energy of 500 $\pm$ 50 keV has been simulated to save resources. 

The results of the simulation can bee seen in Fig. \ref{Graph simulation trajectories}, where it is show the trajectory of the simulated proton beam (left), the comparison between the energy dispersion and temporal dispersion at the source and at the end (right upper and right lower respectively). As we can see, it is possible to obtain extremely monochromatic and narrow pulses at the exit of the device.

\begin{figure}[t]
	\centering
		\includegraphics[width=0.45\textwidth]{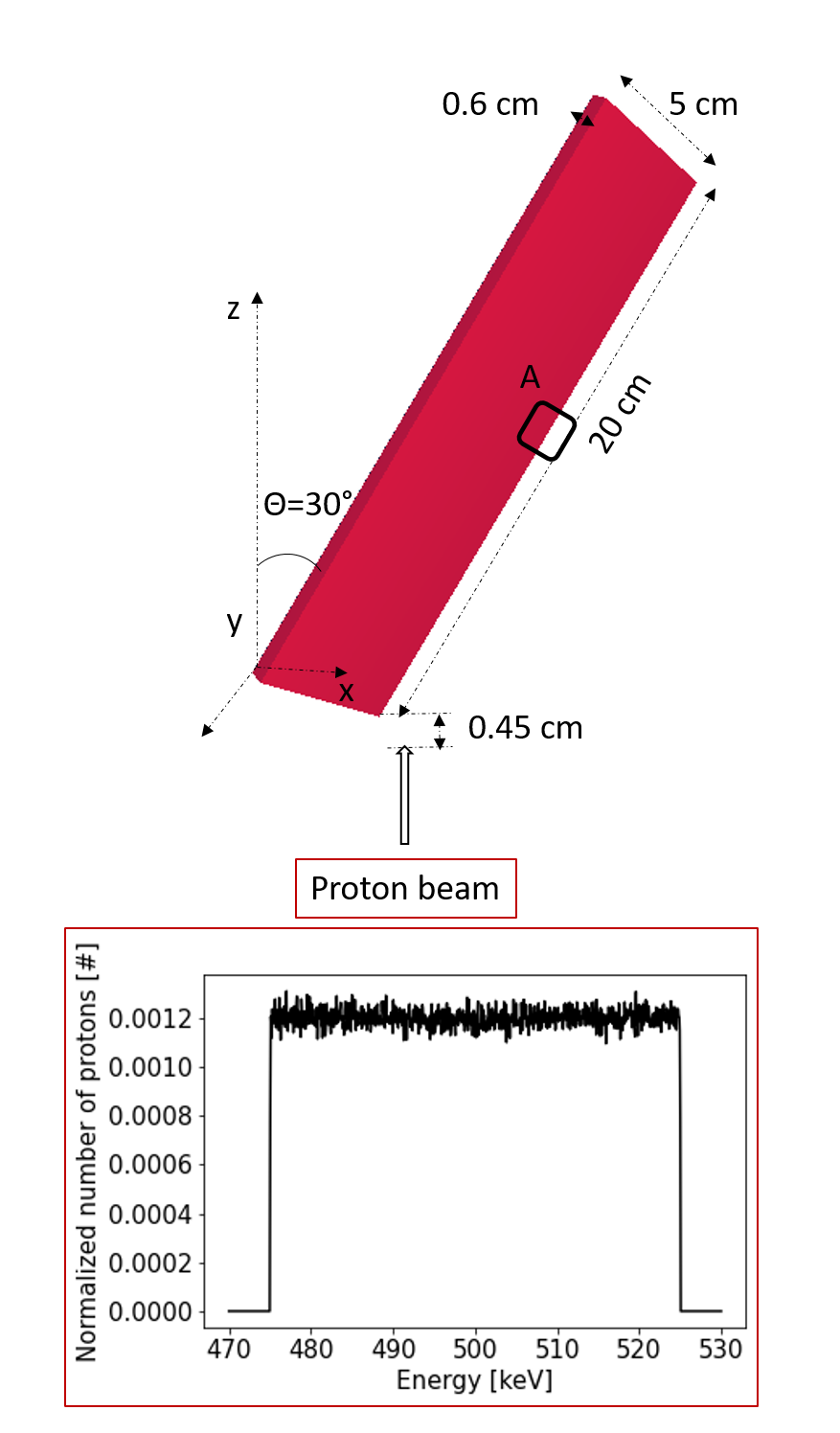}%
	\caption{VisedX intial conditions for the MCNP6 simulation. The A zone is the output region of the proton beam and the inset is the proton energy distribution }
\label{MCNP6_Magnet}
\end{figure}
By way of comparison , trajectories, spot and final energy distribution have been simulated by MCNP6 software (based on Monte Carlo method) which includes routines for tracking charged particles through magnetic fields by direct integration methods \cite{bull2011magnetic}. 

The following parameters have been considered to describe the geometrical set up for MCNP6 simulations: the dimensions of the rectangular cell that forms the dipole are 20 cm length (Z dimension), 5 cm width (X dimension) and 0.6 cm height (Y dimension). The magnet is tilted 30 degrees with respect to the vertical axis (clockwise sense). Furthermore, the energy of the proton source has been set within the range 500 $\pm$ 25 keV ( Fig.\ref{MCNP6_Magnet} inset, represents the proton initial distribution as a function of the energy). Spatially the initial beam represents a Dirac delta without width. In addition, no divergence was considered for these simulations.  

The proton source was placed at 4.5 mm from the closest point of the magnet and the magnetic B-field modulus has been established as 1 Tesla, with direction normal to the magnet surface. The results of the trajectories are obtained by means of FMESH superimposed MESH tally B type FLUX \cite{osti_1169677} that allows to solve the problem over the system geometry. 

\begin{figure}[t]
	\centering
		\includegraphics[width=0.45\textwidth]{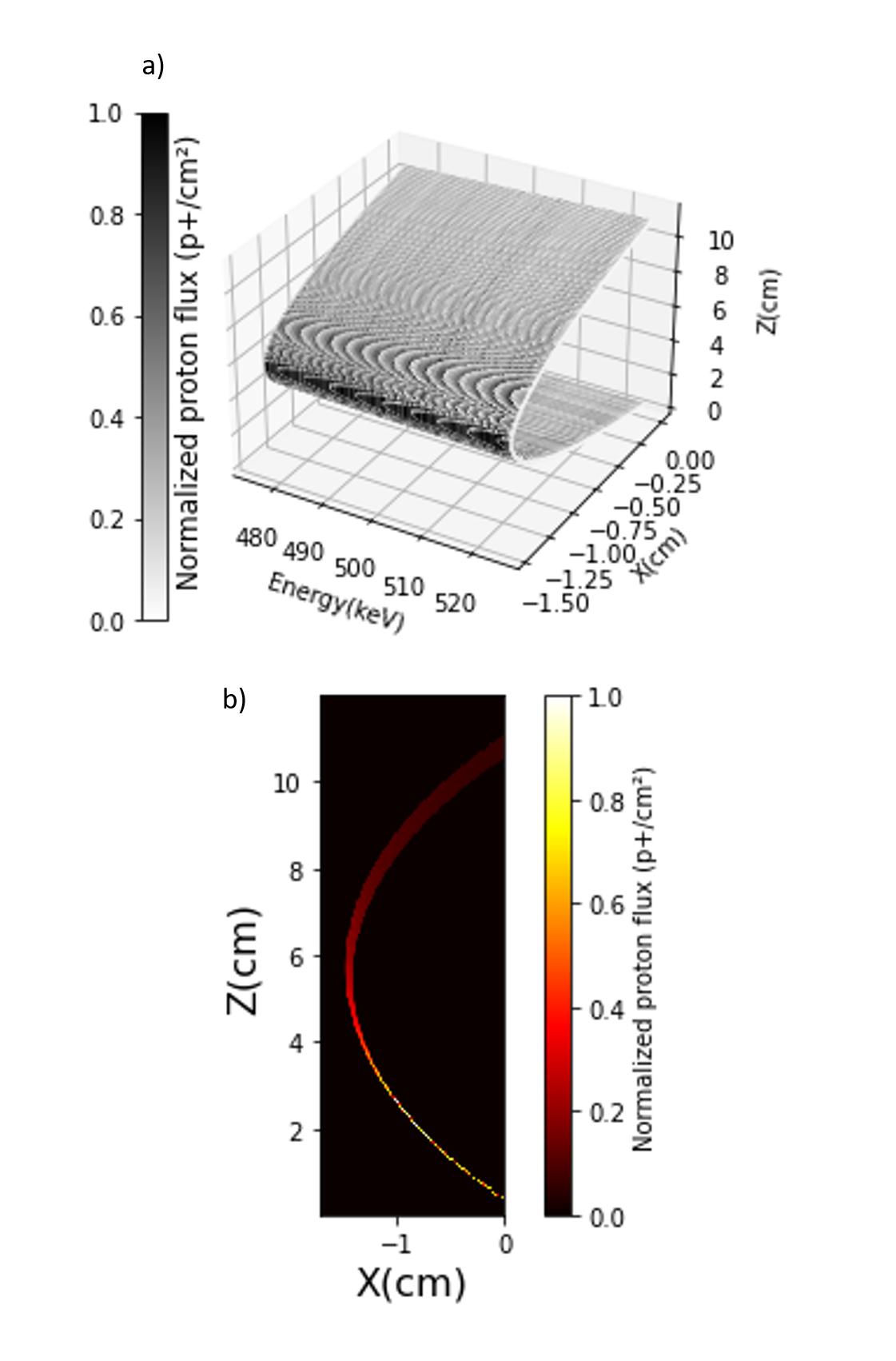}%
	\caption{a) MCNP6 proton trajectories as a function of the energy. b) Projection of the proton flux over the magnet surface}
\label{flux}
\end{figure}

The results are shown in Fig.\ref{flux} a) and b) that represent the trajectories as a function of the energy and the projection of the flux over the magnet surface, respectively. The lower energy protons find the exit at 9.95 $\pm$ 0.02 cm from the input point (at 0.4 $\pm$ 0.02 cm from the magnet bottom corner in the simulation) and the higher energy ones at 10.50 $\pm$ 0.02 cm. These parameters match with expression (7) that results in 10.23 cm which is approximately the midpoint between the two values mentioned above.

\begin{figure}[t]
	\centering
		\includegraphics[width=0.45\textwidth]{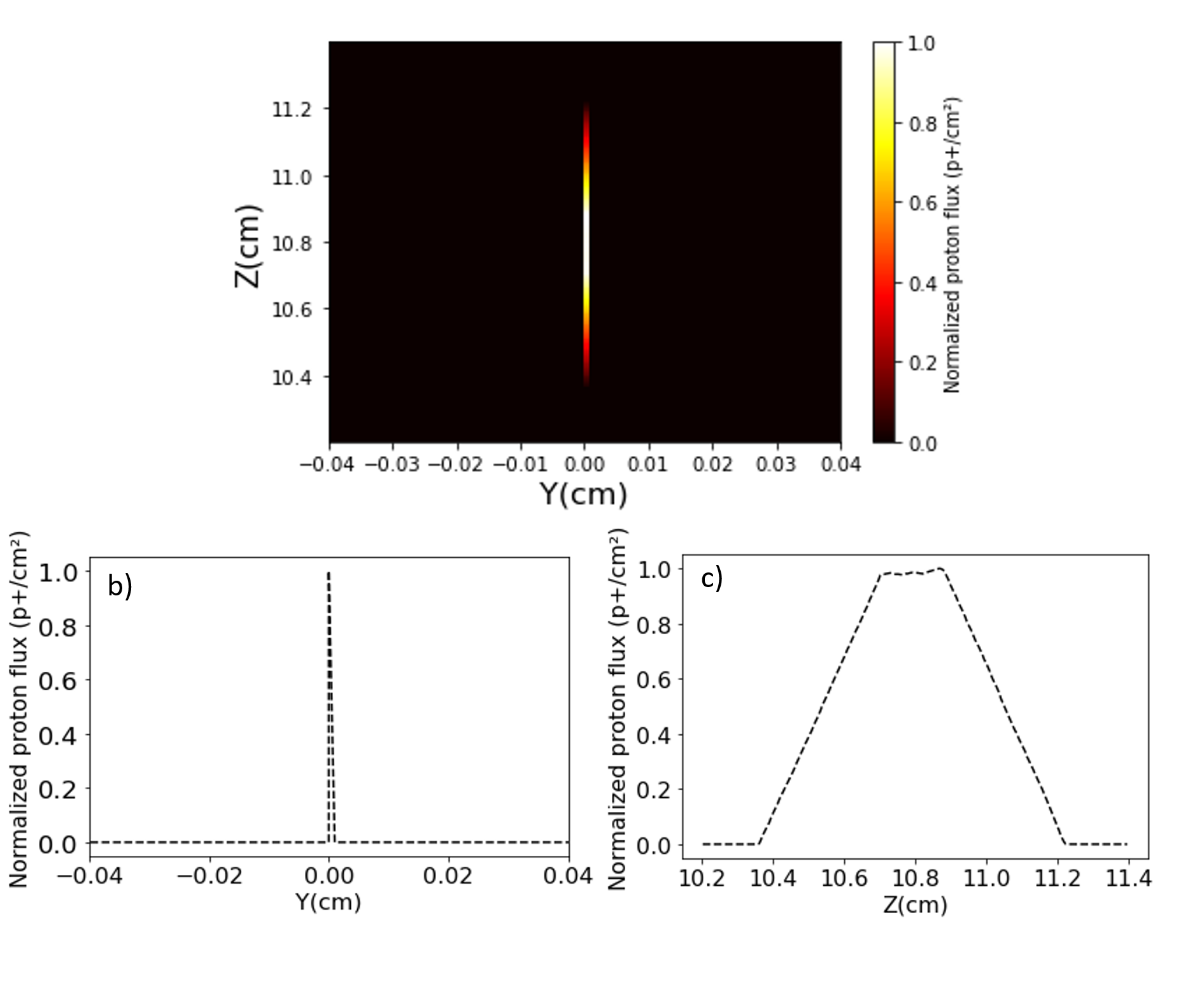}%
	\caption{a) Spot at the exit of the magnet. b) and c) are the vertical and horizontal profiles, respectively}
\label{spot}
\end{figure}

Other outcome of MCNP6 simulations is the spot shape at the exit of the magnet in which, the particle distribution at different energies along the z-axis can be observed (Fig. \ref{spot} a)). Figures (\ref{spot} b) and c)) represent the normalized projections of the spot over the axes of Fig. \ref{spot} a) (profiles). In these simulations it can be observed the extension of the proton beam in Z, while in Y dimension remains sharped. Using the expression (8) with the MCNP6 applied parameters, the value of ($r_{pinh}$) corresponds to 0.26 cm. This value fits with the half of FWHM in Fig.\ref{spot} c) (spot profile projection in Z) making evident the spatial selection of particles as a function of their energy.

(Fig. \ref{spot})

\section{Discussion and conclusions}
A novel and simple design has been proposed for a isochrone magnetic selector capable of angular and energy selection for laser-driven proton pulses by minimizing their temporal duration. The proposed device can finally select proton beams with central energies of hundreds of keV and tens of keV bandwidth and transport them up to the sample maintaining the temporal duration below 100s of ps, which is the required temporal duration to probe micrometeric size laser-driven Warm Dense Matter. A preliminary parametric study is presented by comparing the performances with a previous magnetic selector used in a recent experiment. The comparison shows a consistent and relevant reduction x6 of the final proton pulse duration. A full Monte Carlo simulation has been performed by using the MCNP6 code \cite{mcnpx} consolidating the results. 

\section*{Acknowledgements}
The authors acknowledge the financial support of the Junta de Castilla y Le\'on grant no. CLP263P20.

\bibliographystyle{plain}
\bibliography{references}{}

\end{document}